\documentclass[journal]{IEEEtran}
\IEEEoverridecommandlockouts
\usepackage{amsfonts}
\usepackage{amssymb}
\usepackage[cmex10]{amsmath}
\usepackage{graphicx}
\usepackage{subfigure}
\usepackage{verbatim}
\usepackage{cite}
\usepackage{diagbox}
\usepackage{tabularx,booktabs}
\usepackage{amsthm}
\usepackage{booktabs}

\usepackage{multirow}
\usepackage{xypic}
 \usepackage{enumerate}

\usepackage[ruled,lined]{algorithm2e}
\usepackage{setspace}
\usepackage{stfloats}
\usepackage{fancyhdr}
\usepackage{float}
\usepackage{algorithmic}

\begin{document}
\newcommand{\be}{\begin{equation}}
\newcommand{\ee}{\end{equation}}
\newcommand{\br}{{\mbox{\boldmath{$r$}}}}
\newcommand{\bp}{{\mbox{\boldmath{$p$}}}}

\newcommand{\bn}{{\mbox{\boldmath{$n$}}}}
\newcommand{\balfa}{{\mbox{\boldmath{$\alpha$}}}}
\newcommand{\ba}{\mbox{\boldmath{$a $}}}
\newcommand{\bta}{\mbox{\boldmath{$\beta $}}}
\newcommand{\bg}{\mbox{\boldmath{$g $}}}
\newcommand{\bPsi}{\mbox{\boldmath{$\Psi $}}}
\newcommand{\bpsi}{\mbox{\boldmath{$\psi $}}}
\newcommand{\bsigma}{\mbox{\boldmath{ $\Sigma $}}}
\newcommand{\bpi}{\mbox{\boldmath{ $\pi $}}}
\newcommand{\bGamma}{{\bf \Gamma }}
\newcommand{\bA}{{\bf A }}
\newcommand{\bII}{{\bf I }}
\newcommand{\bP}{{\bf P }}
\newcommand{\bX}{{\bf X }}
\newcommand{\bI}{{\bf I }}
\newcommand{\bR}{{\bf R }}
\newcommand{\bff}{{\mathbf{f}}}
\newcommand{\bZ}{{\bf Z }}
\newcommand{\bz}{{\bf z }}
\newcommand{\bx}{{\mathbf{x}}}
\newcommand{\bM}{{\bf M}}
\newcommand{\bW}{{\bf W}}
\newcommand{\bU}{{\bf U}}
\newcommand{\bD}{{\bf D}}
\newcommand{\bJ}{{\bf J}}
\newcommand{\bH}{{\bf H}}
\newcommand{\bK}{{\bf K}}
\newcommand{\bm}{{\bf m}}
\newcommand{\bN}{{\bf N}}
\newcommand{\bC}{{\bf C}}
\newcommand{\bL}{{\bf L}}
\newcommand{\bF}{{\bf F}}
\newcommand{\bv}{{\bf v}}
\newcommand{\bSigma}{{\bf \Sigma}}
\newcommand{\bS}{{\bf S}}
\newcommand{\bs}{{\bf s}}
\newcommand{\bO}{{\bf O}}
\newcommand{\bQ}{{\bf Q}}
\newcommand{\btr}{{\mbox{\boldmath{$tr$}}}}
\newcommand{\bNSCM}{{\bf NSCM}}
\newcommand{\barg}{{\bf arg}}
\newcommand{\bmax}{{\bf max}}
\newcommand{\test}{\mbox{$
\begin{array}{c}
\stackrel{ \stackrel{\textstyle H_1}{\textstyle >} } { \stackrel{\textstyle <}{\textstyle H_0} }
\end{array}
$}}

\newtheorem{Def}{Definition}
\newtheorem{Pro}{Proposition}
\newtheorem{Rem}{Remark}
\newtheorem{Lem}{Lemma}
\title{Distributed Fusion of Labeled Multi-Object Densities Via Label Spaces Matching}
\author{
Bailu Wang, Wei Yi, Suqi Li, Lingjiang Kong and Xiaobo Yang\\
\IEEEauthorblockA{{University of Electronic Science and Technology of China, School of Electronic Engineering, Chengdu City, China} \\
{Email: kussoyi@gmail.com}}}

\maketitle

\begin{abstract}
In this paper, we address the problem of the distributed multi-target tracking with labeled set filters in the framework of Generalized Covariance Intersection (GCI). Our analyses show that the label space mismatching (LS-DM) phenomenon, which means the same realization drawn from label spaces of different sensors does not have the same implication, is quite common in practical scenarios and may bring serious problems. Our contributions are two-fold. Firstly, we provide a principled mathematical definition of ``label spaces matching (LS-DM)'' based on information divergence, which is also referred to as LS-M criterion.  Then, to handle the LS-DM, we propose a novel two-step distributed fusion algorithm, named as GCI fusion via label spaces matching (GCI-LSM). The first step is  to match the label spaces from different sensors. To this end,  we  build  a ranked assignment problem and design a cost function consistent with LS-M criterion to seek the optimal solution of matching correspondence between label spaces of different sensors. The second step is to perform the GCI fusion on the matched label space.  We also derive the GCI fusion with generic labeled multi-object (LMO) densities based on LS-M, which is the foundation of  labeled distributed fusion algorithms. Simulation results for Gaussian mixture implementation highlight the performance of the proposed GCI-LSM algorithm in two different tracking scenarios.
\end{abstract}

%

\section{Introduction}
 Compared with centralized multi-object tracking methods, distributed multi-sensor multi-object tracking (DMMT) methods generally benefit from lower communication cost and higher fault tolerance. As such, they have increasingly attracted interest from tracing community.  When the correlations between the estimates from different sensors are not known, devising DMMT solutions becomes particularly challenging. The optimal fusion to this problem was developed in \cite{CY-Chong}, but the computational cost of calculating the common information can make the solution intractable in practical applications. An alternative is to use suboptimal fusion technique, namely, Generalized Covariance Intersection (GCI) or exponential mixture Densities (EMDs) \cite{EMD-Julier} pioneered by Mahler \cite{Mahler-1}. The highlight of GCI is that it is capable to fuse  both Gaussian and non-Gaussian formed multi-object distributions from different sensors with completely unknown correlation.

Based on GCI fusion rule, distributed fusion with the probability hypothesis density \cite{Ristic_PHD,PHD-Vo} (PHD)/cardinalized PHD \cite{Franken, Vo-CPHD} and multi-Bernoulli \cite{MeMBer_Vo2,MeMber_Vo3,Reza_ground_target,Reza_sensor_control_letter,Reza_sensor_control_AES,Reza_audio_visual,Reza_visual_tracking} filters  has  been explored in \cite{Mahler-1,Clark,Uney-2,Battistelli, Mehmet, MeMBer_Mahler,GCI-MB}.
However, the aforementioned filters on one hand are not multi-object trackers as target states are indistinguishable, and on the other hand are almost not the closed-form solution to the optimal Bayssian filter even though  a special observation model, i.e., standard observation model  \cite{MeMBer_Mahler} is assumed.
Recently, the notion of labeled random finite set (RFS) is introduced to address target trajectories and their uniqueness in\cite{LMB_Vo, LMB_Vo2, delta_GLMB, LRFS_Bear_Vo, LRFS_Papi_Kim, Fantacci-BN, GLMB_Papi_Vo}. Vo \textit{et al.} proposed a class of  generalized labeled multi-Bernoulli (GLMB) \footnote{GLMB distribution is also simply named as Vo-Vo distribution by Malher in his book \cite{refr:tracking-2} first time. }
 densities which is a conjugate prior and also closed under the Chapman-Kolmogorov equation for the standard observation model in Bayesian inference. Moreover, the relevant stronger results, $\delta$-GLMB filter, which can be directly used to multi-target tracking, can not only produce trajectories formally but also outperform the aforementioned filters. 
Except for the standard observation model, the labeled set filter  also has achieved some good results for generic observation model. In  \cite{GLMB_Papi_Vo}, Papi \textit{et al.} proposed a $\delta$-GLMB density approximation of the labeled multi-object (LMO) density  and  developed an efficient $\delta$-GLMB filter for the generic observation model. \cite{GLMB_Papi_Vo} also provides a further detailed expression for the universal LMO density, which is the product of the joint existence probability of the label set and the joint probability density of states conditional on their corresponding labels.

Due to the advantages of labeled  set filters, it is meaningful to  investigate their generalization to the distributed environment. In \cite{Fantacci-BT}, Fantacci \textit{et al.} derived the closed-form solutions of GCI fusion with marginalized $\delta$-GLMB (M$\delta$-GLMB) and labeled multi-Bernoulli (LMB) posteriors, and highlight the performance of the relevant DMMT algorithms  based on the assumption that different sensors share the same label space. However, our  analyses show that this assumption is  hard to be satisfied in many real world applications. In other word, the label spaces of each sensors always mismatch in the sense that the same realization drawn from label spaces of different sensors does not have the same implication in practical scenarios, which is  referred to as  ``label space mismatching (LS-DM)''\footnote{The letter ``D'' in the abbreviation of label space mismatching means double ``M'', i.e., ``miss'' and ``matching''.}. When LS-DM happens, the direct fusion with the labeled posteriors from different sensors will exhibit a counterintuitive behavior: the fusion will be performed between objects with different labels, making the fusion performance  poor. Therefore, there is a lack of robustness in practice if one perform fusion with labeled posteriors from different sensors directly.

To get rid of the bad influences of  LS-DM, two promising thoughts can be employed: one is to perform GCI fusion with unlabeled version of posteriors from different sensors, which is firstly proposed in \cite{GCI-GMB}, and  the other is to match the label spaces of different sensors and then perform  GCI fusion  on the matched label space. This paper focuses on the latter and our contributions are two-fold:
 \begin{enumerate}[i)]
 \item
 \textit{We provide a principled mathematical definition for ``label space matching'' based on  information divergence}.
This definition also provides a criterion to judge whether the label spaces are matching or not.
Moreover to make this criterion have practicality, we derive the specified expression of set marginal density for single-object case.
 \item
 \textit{We proposed a two-step distributed fusion algorithm, namely, GCI fusion with LMO densities via label spaces matching (GCI-LSM for short).} First step is to match the label spaces from different sensors. To this end, the ranked assignment problem is built to seek the optimal solution of the matching correspondence with the cost function based on LS-M criterion. Then, perform GCI fusion with LMO densities on matched label space. In addition, we derive the GCI fusion with the generic LMO density based on the assumption that label spaces are matching, which is the foundation of many labeled DMMT..
\end{enumerate}

In numerical results, the performance of the proposed fusion algorithm with Gaussian mixture (GM) implementation is verified.
\section{Background}
\subsection{Notation}
In this paper, we inhere the convention that single-target states are denoted by the small letter ``x'', e.g., $x,\bx$ and the multi-target states are denoted by capital letter ``X'', e.g., $X,\bX$.  To distinguish labeled states and distributions from the unlabeled ones, bold face letters are adopted for the labeled ones, e.g., $\bx$, $\bX$, $\bpi$. Observations generated by single-target states are denoted by the small letter ``z'', i.e., $z$, and the multi-target observations are denoted by capital letter ``Z'', i.e., $Z$. Moreover, blackboard bold letters represent spaces, e.g., the state space is represented by $\mathbb{X}$, the label space by $\mathbb{L}$, and the observation space  by $\mathbb{Z}$. The collection of all finite sets of $\mathbb{X}$ is denoted by $\mathcal{F}(\mathbb{X})$ and $\mathcal{F}_n(\mathbb{X})$ denotes all finite subsets with $n$ elements.

The labeled single target state $\bx$ is constructed by augmenting a state $x\in\mathbb{X}$ with an label $\ell\in\mathbb{L}$. The labels are usually drawn form a discrete label space, $\mathbb{L}=\{\alpha_i:i\in\mathbb{N}\}$, where all $\alpha_i$ are distinct and the index space $\mathbb{N}$ is the set of positive integers.

The multi-target state $X$, the labeled multi-target state $\bX$ and the multi-target observation $Z$ are modelled by the finite set of single-target states, the finite set of labeled single-target states, and the finite set of  observations generated by single-target states, respectively, i.e.,
\begin{equation}\label{multi-target state}
\begin{split}
  X=&\{x_1,\cdots,x_n\}\subset\mathbb{X}\\
  \bX=&\{\bx_1,\cdots,\bx_n\}\subset\mathbb{X}\times\mathbb{L}\\
  Z=&\{z_1,\cdots,z_m\}\subset\mathbb{Z}
\end{split}
\end{equation}

We use the multi-object exponential notation
\begin{equation}\label{multi-object exponential notation }
  h^{X}\triangleq\prod_{x\in X}h(x)
\end{equation}
for real-valued function $h$, with $h^\emptyset=1$ by convention.
To admit arbitrary arguments like sets, vectors and integers, the generalized Kronecker delta function is given by
\begin{equation}\label{delta}
  \delta_Y(X)\triangleq\left\{\begin{array}{l}
\!\!1, \mbox{if}\,\,\,X=Y\\
\!\!0, \mbox{otherwise}
\end{array}\right.
\end{equation}
and the inclusion function is given by
\begin{equation}\label{inclusion function}
  1_Y(X)\triangleq\left\{\begin{array}{l}
\!\!1, \mbox{if} X\subseteq Y\\
\!\!0, \mbox{otherwise}
\end{array}\right.
\end{equation}
If $X$ is a singleton, i.e., $X=\{x\}$, the notation $1_Y(X)$ is used instead of $1_Y(\{x\})$.

 Also notice that the labeled multi-target state is an RFS on $\mathbb{X}\times\mathbb{L}$ with distinct labels. The set of labels of an labeled RFS $\mathbb{X}$ is given by $\mathcal{L}(\bX)=\{\mathcal{L}(\bx):\bx\in\bX\}$, where $\mathcal{L}:\mathbb{X}\times\mathbb{L}\rightarrow\mathbb{L}$ is the projection defined by $\mathcal{L}((x,\ell))=\ell$.
 The distinct label indicator
\begin{equation}
  \bigtriangleup(\bX)=\delta_{|\bX|}(|\mathcal{L}(\bX)|)
\end{equation}
\subsection{Labeled Multi-Object Density}
For an arbitrary labeled RFS, its LMO density can be represented as the expression given in Lemma 1
  \cite{GLMB_Papi_Vo}.
 \begin{Lem} Given an labeled multi-object density $\bpi$ on $\mathcal{F}(\mathbb{X}\times\mathbb{L})$, and for any positive integer $n$, we define the joint existence probability of the label set $\{\ell_1,\ell_2,\cdots,\ell_n\}$ by
\begin{equation}\label{joint existence probability}
  w(\{\ell_1,\cdots,\ell_n\})\!=\!\int\!\! \bpi(\{(x_1,\ell_1),\!\cdots,\!(x_n,\ell_n)\})d(x_1,\!\cdots\!,x_n)
\end{equation}
and the joint probability density on $\mathbb{X}^n$ of the states $x_1,\cdots,x_n$ conditional on their corresponding labels $\ell_1,\cdots,\ell_n$ by
\begin{equation}\label{joint probability density}
  P(\{(x_1,\ell_1),\cdots,(x_n,\ell_n)\})=\frac{\bpi(\{(x_1,\ell_1),\cdots,(x_n,\ell_n)\})}{w(\{\ell_n,\cdots,\ell_n\})}
\end{equation}
Thus, the LMO density can be expressed as
\begin{equation}\label{factorized}
  \bpi(\mathbf{X})=w(\mathcal{L}(\mathbf{X}))P(\mathbf{X}).
\end{equation}
\end{Lem}
\subsection{Multi-target Bayesian filter}
Finite Set Statistics (FISST) proposed by Mahler, has provided a rigorous and elegant mathematical framework for the multi-target detection,
tracking and classification problem in an unified Bayesian paradigm.

In the FISST framework, the optimal multi-target Bayesian
filter \footnote[1]{Note that the multi-object Bayesian filter  in (\ref{Optimal-prediction}) and (\ref{Optimal-update}) is also appropriate for the labeled set posterior, and the labeled set integrals defined as \cite{LMB_Vo} are involving.} propagates RFS based posterior density $\pi_{k}(X_k|Z^{1:k})$ conditioned on the sets of observations up to time $k$, $Z^{1:k}$, in time with
the following recursion  \cite{MeMBer_Mahler}:
\begin{align}\label{Optimal-prediction}
\begin{split}
&\pi_{k|k-1}(X_k|Z^{1:k-1})\\&=\int f_{k|k-1}(X_k|X_{k-1})\pi_{k-1}(X_{k-1}|Z^{1:k-1})\delta X_{k-1},
\end{split}
\end{align}
\begin{align}\label{Optimal-update}
\begin{split}
\pi_{k}(X_k|Z^{1:k})=\frac{g_k(Z_k|X_k)\pi_{k|k-1}(X_k|Z^{1:k-1})}{\int g_k(Z_k|X_k)\pi_{k|k-1}(X_k|Z^{1:k-1})\delta X_{k}}
\end{split}
\end{align}
where $f_{k|k-1}(X_k|X_{k-1})$ is the multi-target Markov transition function and $g_k(Z_k|X_k)$ is the multi-target likelihood function of $Z_k$, and $\int\cdot\delta X$ denotes the set integral \cite{MeMBer_Mahler} defined by
\begin{equation}\label{set integral}
  \int\! f(X)\delta X\!=\sum_{n=0}^\infty \frac{1}{n!}\int\! f(\{x_1,\cdots,x_n\})dx_1\cdots dx_n.
\end{equation}
\subsection{GCI Fusion Rule}
The GCI was proposed
by Mahler specifically to extend FISST to distributed
environments [9].
Consider two nodes $1$ and $2$ in the sensor network. At time $k$, each nodes maintain its own local posteriors $\pi_{1}(X|Z_{1})$ and $\pi_{2}(X|Z_{2})$ which are both the RFS based densities. Under the GCI \footnote[2]{Note that GCI fusion rule in (\ref{G-CI}) is also appropriate for the labeled set posterior, and the labeled set integrals defined as \cite{LMB_Vo} are involving.} proposed by Mahler, the fused distribution is the geometric mean, or the exponential mixture of the local posteriors \cite{Mahler-1},
\begin{align}\label{G-CI}
\begin{split}
\pi_{\omega}(X|Z_{1},Z_{2})=\frac{\pi_{1}(X|Z_{1})^{\omega_1}\pi_{2}(X|Z_{2})^{\omega_2}}
                                              {\int \pi_{1}(X|Z_{1})^{\omega_1}\pi_{2}(X|Z_{2})^{\omega_2}\delta X}
\end{split}
\end{align}
where $\omega_1$, $\omega_2$ ($\omega_1+\omega_2=1$) are the parameters determining the relative fusion weight of each distributions.

Eq.~(\ref{G-CI}) is derived by following that the distribution that minimizes the weighted sum of its Kullback-Leibler divergence (KLD) with respect to a given set of distributions is an EMD \cite{Mahler-1},
\begin{equation}\label{EMD}
  \pi_\omega=\arg \min_\pi(\omega_1D_{\emph{KL}}(\pi\parallel \pi_1)+\omega_2 D_{\emph{KL}}(\pi\parallel \pi_2))
\end{equation}
where $D_{\emph{KL}}$ denotes the KLD with
\begin{equation}\label{KLD}
\begin{split}
 D_{\emph{KL}}(f||g)\triangleq\int f(X)\log{\frac{f(X)}{g(X)}}dX.
  \end{split}
\end{equation}
Note the integral in (\ref{KLD}) must be interpreted as a set integral.
\section{Label Space Mismatching Phenomenon and
 Promising Solutions}
The GCI formula in (\ref{G-CI}) is generally computationally intractable for  the set integrals need to integrate over all joint  state spaces, considering each cardinality (number of objects).     Fortunately, it is tractable to derive the closed-form solutions for GCI fusion with many simplistic labeled densities including LMB and M$\delta$-GLMB densities, which can simplify
the G-CI formula largely. However, these closed-form solutions are derived based on the assumption that  the label spaces of different local  labeled set filters are matching, and  this assumption is really harsh  in practice making these solutions   restrictive in realworld DMMT.

In this section,  we firstly analyze the causes of \textit{label spaces mismatching} (LS-DM) phenomenon in terms of two popular birth procedures, and then provide two novel methods to solve this challenge problem.
\subsection{LS-DM Phenomenon}
The essential meaning  of LS-DM  is that the same realization drawn from label spaces of different sensors does not have the same implication. The underlying implication LS-DM is that the posterior spatial distributions of the same object in different sensors only have  a tiny discrepancy.  This phenomenon is quite  common in labeled DMMT. It may originate from any time steps during the recursion of multi-object filtering and fusing, and  will last during the subsequent time steps in many cases.  Naturally, the birth procedure has a  decisive influence on the matching of label spaces of different sensors.  Hence, in the following, we analyze the causes of  LS-DM in terms of two popular birth procedures.
\begin{enumerate}[(1)]
\item \textit{Adaptive birth procedure (ABP) \cite{LMB_Vo2}}. This birth procedure is widely used, in which new-born targets are based on the observations not associated to the persisting targets. Due to the randomness of the observations, it is really difficult to guarantee that the same births of different local set filters are labeled using the same object label.   In addition, the observation sets provided by different sensors incorporating noisy observations of objects, stochastic miss-detections,  and  stochastic clutters are also contributed to the LS-DM of persisting objects. For instance, if sensor 1 loses a object due to miss-detection and  re-initiate it later, while sensor 2 keep locking on this object always,  then the mismatching of the label of this object will arise.

\item \textit{Priori knowledge based on birth procedure (PBP)\cite{MeMBer_Vo2}}. This birth procedure is often used in some well-known scenarios with the priori of positions of  object births, e.g., entrance of marketplace, airport etc.  Generally, the object label has two dimension, $\ell=(k,i)$, where $k$ is the time of birth, and $i$ is a unique index to distinguish objects. The priori of the born positions for PBP can provide reference for the object index $\ell$,  but contribute little to the birth time $k$, hence there still exists a chance of mismatching for the births  since it is easily effected by  the uncertain of measurement noise, clutter and the variance of the prior  of the born position. In addition, due to that persisting objects may be wrongly dominated by clutters, or  be truncated due to miss-detection in the following time steps, the LS-DM for persisting objects also happen sometimes.
\end{enumerate}

 The above analyses suggest that for  both ABP and PBP, it is  difficult to ensure different sensors share the same label space. Note that   PBP suffer less from LS-MM than ABP for it can use the prior information as a reference.
In a word, to ensure the matching of label spaces of each sensors, an ideal detecting environment, in which each sensor dose not have miss-detections and clutters, and the estimate accuracy of each sensor is enough high, is required.


\subsection{Promising Solutions for LS-DM}
To break away the bad influence of LS-MM,  we propose two solutions:
\begin{enumerate}[$\bullet$]
\item The first method \cite{GCI-GMB} is that the GCI fusion is performed on unlabeled state space via transforming the labeled RFS densities to their unlabeled versions. Therefore, this fusion method has robustness. For the GLMB family, we had proved that their unlabeled versions is the generalized multi-Bernoulli  (GMB) distributions \cite{GCI-GMB}, and the GCI fusion with GMB distributions  (GCI-GMB) is also proposed in \cite{GCI-GMB}.
\item The second method is that  firstly  match the label spaces from different sensors, then perform GCI fusion with labeled densities on macthed labeled state space as shown in Fig 1.  This approach is referred to as GCI fusion with label space matching (GCI-LSM).
\end{enumerate}

This paper mainly focuses on the GCI-LSM fusion method.
\begin{figure}[htpb]\label{fig_GCI_LSM}
\centering
\includegraphics[width=3.2in]{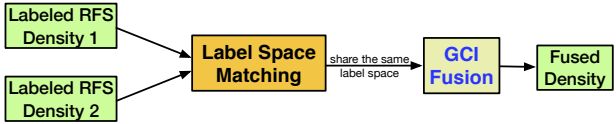}
\caption{{ GCI-LSM: the label spaces of different sensors are matched through some means each time firstly, then perform GCI fusion on labeled state space.}}
\label{two_mehtod}
\end{figure}
\section{GCI Fusion  via Label Space Matching}
Based on the second solution of LS-DM, label spaces matching and the closed-form solution of GCI fusion with LMOs are the two key points need to be addressed. To clearly describe the concept of label space matching (LS-M), we firstly give the mathematical definition of label spaces matching based on information divergence, also referred to as LS-M criterion, which is the foundation of GCI-LMS fusion method.  Then by solving the built ranked assignment problem about the matching relationship of objects between different sensors, we then get matched  label spaces. Finally with this condition different label spaces are matched,  the closed-form solution of GCI fusion with universal LMO densities is derived.
\subsection{The Mathematical Description of Label Space Matching}
In order to clearly describe ``label space matching'',  we formulate it in a rigorous mathematical model shown in Definition 1. Definition 1 also provides a criterion  to judge whether the label spaces from different sensors are matching or not, and we  call this criterion  as \textit{LS-M criterion}.
\begin{Def}
Consider a scenario that sensor 1 and sensor 2 observe the same spatial region. Suppose that  $\bpi_1(\cdot)$ and $\bpi_2(\cdot)$  with state space $\mathbb{X}$ and label space $\mathbb{L}_i$ are the multi-object posteriors of sensor 1 and sensor 2 respectively. The RFS $\bPsi_i, i=1,2$ with the probability density $\bpi_i(\cdot)$ can be represented as the union
\begin{equation}\label{union}
\bPsi_i=\biguplus_{\ell\in\mathbb{L}} \bpsi_i^{(\ell)}
\end{equation}
where  $\bpsi_i^{(\ell)}$ with the state space $\mathbb{X}\times \mathbb{\{\ell\}}$ is the random finite subset of $\bPsi$.

Then $\mathbb{L}_1$ and $\mathbb{L}_2$ are said to be matching only if
\begin{equation}\label{LS-M}
\mathbb{L}_1=\mathbb{L}_2\,\,\,\,\mbox{and}\,\,\,\,\forall~\ell\in \mathbb{L}_{1}, \  D(\bpi_1^{(\ell)}||\bpi_2^{(\ell)})\leq\Gamma_m
\end{equation}
where $\bpi_s^{(\ell)}(\cdot)$ is the probability density of $\bpsi_s^{(\ell)}$, $\ell\in \mathbb{L}_s, s=1,2$,  $\Gamma_m$ is a given threshold, and $D(\bpi_1^{(\ell)}(\cdot)||\bpi_2^{(\ell)}(\cdot))$ describes the ``distance'' between two distributions.
\end{Def}

The condition that $\mathbb{L}_1=\mathbb{L}_2$ demands that both the cardinalities and each elements of the label spaces of sensor 1 and sensor 2  have the same values. Note $\bpsi_s^{(\ell)}$ is the random finite subset related the object with label $\ell$, and $\bpi_s^{(\ell)}$ is the probability density for object $\ell$ in sensor $s$.  Hence, the condition $ D(\bpi_1^{(\ell)}||\bpi_2^{(\ell)})<\Gamma_m$ demands that the densities of object $\ell$ in sensors 1 and 2 have a slight difference, which ensures that the object label $\ell$ of sensors 1 and 2 are matching. As $\bpi_s^{(\ell)}$ incorporates all the statistical information about object $\ell$, and thus it is reasonable to judge the matching relationship of label $\ell$ based on its probability densities.    In a word, to match the label spaces $\mathbb{L}_1$ and $\mathbb{L}_2$, one-to-one object matching constrain should be satisfied between different sensors.
\begin{Rem}
The parameter $\Gamma_m$ in (\ref{LS-M}) is  a given threshold, and the slighter its values is, the harsher the LS-M criterion is. In ideal case, $\Gamma_0$ which means the objects of different sensors share the same density.
 The ``distance'' $D(f||g)$ in (\ref{LS-M}) usually chooses information-based divergences including KLD, R$\acute{e}$nyi, Csisz$\acute{a}$r-Morimoto (AliSilvey), and Cauchy-Schwarz, etc., to measure the similarlies/differences of different multi-object density.
\end{Rem}

%

 A key point of using the LS-M criterion  is to compute the probability density $\bpi_s^{(\ell)}(\cdot)$ from the global multi-object density $\bpi_s(\cdot)$. $\bpi_s^{(\ell)}(\cdot)$ is also called as set marginal density of $\bpi_s^{(\ell)}(\cdot)$ with respective to $\bpi_s(\cdot)$. In \cite{refr:JMB}, we preliminarily give the concept of the set marginal density as
shown in Definition 3, its generalized computing method as
shown in Lemma 1,  and  its specified computing method for joint
multi-Bernoulli RFS. In \cite{Li_1}, the set marginal density is extened to labeled multi-object density.  In this section, we derive the specified expression for the set marginal density of  labeled random finite subset $\bpi_s^{(\ell)}$ with respective to $\bpi_s$  in Proposition 1. Proposition 1 guarantees the practicability of the LS-M criterion.

\begin{Def}  Let $\Psi$ be an RFS. Then for any random finite subset of $\Psi$, denoted by $\psi$, its set density function $f_\psi(X)$, is called set marginal density of  $\psi$ with respect to $\Psi$.
\end{Def}
\begin{Lem}
 Let $\bPsi$ be an RFS. Then for any random finite subset of $\bPsi$, denoted by $\psi$, its set marginal density of $\psi$ with respect to $\bPsi$, denoted by $f_{\psi}(X)$ can be derived as
\begin{equation}\label{the_marginal_density_of_an_RFS}
 f_{\psi}(X)=\frac{\delta\Pr(\psi\subseteq S,\Psi/\psi\subseteq \mathbb{X})}{\delta X}\bigg{|}_{S=\emptyset}
\end{equation}
where $\Psi/\psi
\triangleq\{\bx|\bx\in\Psi\,\mbox{and}\,\bx\notin\psi\}$ and ``$\delta/\delta X$'' denotes a set derivative \cite{MeMBer_Mahler}.
\end{Lem}

\begin{Rem}
Eq. (\ref{the_marginal_density_of_an_RFS}) (Lemma 2) makes us convenient get a set marginal density of a random finite subset of an labeled RFS $\Psi$. Indeed, the local statistical properties of an labeled RFS can be learned by the set marginal density. Also, the relations of label spaces or correlations among different RFSs densities can be known via analyzing the relevances of their corresponding set marginal densities.
\end{Rem}

According Proposition 2 in \cite{Li_1}, the set marginal density with single object space is derived in the following,
\begin{Pro}\label{marginal_density}
Given  $\bpi_{s}(\cdot)=w_s(\mathcal{L}(\cdot))P_s(\cdot)$ be the multi-object posterior of sensor $s$.
For any $\ell \in \mathbb{L}_s$, the set marginal density of its corresponding subset $\bpsi_s^{(\ell)}$ on space $\mathbb{X}\times \{\ell\}$  is an labeled Bernoulli distribution with parameters $\bpi_s^{(\ell)}=\{(r_s^{(\ell)},p_s^{(\ell)})\}$ are shown as
\begin{align}
r_s^{(\ell)}&=\sum_{I\in \mathcal{F}(\mathbb{L}_s)}1_{I}(\ell)w_s(\cup I),\\
p_s^{(\ell)}&=\frac{1}{r^{(\ell)}} \sum_{I\in \mathcal{F}(\mathbb{L}_s)}1_{I}(\ell)w_s(\cup I) p_{I-\{\ell\}}(\{x,\ell\})
\end{align}
where
\begin{align}
&p_{\{\ell_1,\cdots,\ell\}}(\{x,\ell\})\\&=\int P_s(\{(x,\ell),(x_1,\ell_1),\cdots,(x_n,\ell_n)\})d x_1,\cdots, d x_n.
\end{align}
\end{Pro}
\begin{Rem}
Proposition 1 indicates  that the set marginal density of each $\bpsi_s^{(\ell)}, \ell\in\mathbb{L}_s$  is an labeled Bernoulli distribution. The class of   Bernoulli  densities own a congenital advantage that it can get  tractable results for information divergence generally  making the computation of  $ D(\bpi_1^{(\ell)}||\bpi_2^{(\ell)})<\Gamma_m$ simplistic, thus enhance the practicability of the LS-M criterion largely.
%
%
\end{Rem}

\subsection{Label Space Matching via Ranked Assignment Problem}
Section III-B showed that LS-DM phenomenon is quite common in practical scenarios. Actually, for different sensors observing the same spatial region, the tracks of a sensor have only one  definite correspondence with the tracks of another sensor in ideal case, consistent with $\Gamma_m=0$ in (\ref{LS-M}). However, due to the influence of the stochastic noise, there exist great uncertainty for the matching correspondence.  Then the problem of seeking the solution of matching correspondence is essentially a optimization problem.

 In this section, we firstly provide the mathematical representation of the matching correspondence between different sensors using a mapping function. Then based on the LS-M criterion given in Definition 1, we build a ranked assignment problem and design a principle cost function to seek the solution of optimal matching correspondence, where the information divergence employs the  R\'{e}nyi Divergence (RD) which is the generalized form of the KLD  with the free parameter $\alpha\rightarrow 1$.
\begin{Def}
A fusion map is a function $\tau: \mathbb{L}_1\rightarrow \{0\}\cup\mathbb{L}_2$ such that  $\tau(i)=\tau(i^{'})>0$ implies $i=i^{'}$. The set  of all such fusion maps is called the fusion map space denoted by $\mathcal{T}$.
\end{Def}

Each fusion map $\tau\in\mathcal{T}$ describes one possible (hypothesis) matching relationship of different label spaces,
\begin{align}\label{Mapping}
\begin{split}
\{(\ell,\tau(\ell))\}_{\ell\in\mathbb{L}_1}
\end{split}
\end{align}
and the number of fusion maps grows exponentially with the number of objects.

Due to the uncertainty of $\tau^{*}$,  we need to seek the optimal estimation of  $\tau^{*}$ in order to perform GCI fusion on matching label space. This can be accomplished by solving the following ranked assignment problem.

Enumerating $\mathbb{L}_1$ and $\mathbb{L}_2$, each fusion map $\tau\in\mathcal{T}(\mathbb{L}_1)$ can be represented by an $\left|\mathbb{L}_1\right|\times\left|\mathbb{L}_2\right|$ assignment matrix ${S}$ consisting of $0$ or $1$ entries with every row and column summing to either 1 or 0. For $i\in\{1,\ldots,\left|\mathbb{L}_1\right|\}$, $j\in\{1,\ldots,\left|\mathbb{L}_2\right|\}$, $S_{i,j}=1$ if and only if track $\ell_1^{(i)}$  of sensor 1 is assigned to track $\ell_2^{(j)}$ of sensor 2, i.e. $\tau(\ell_1^{(i)})=\ell_2^{(j)}$. An all-zero row $i$ means that track $\ell_1^{(i)}$ of
sensor 1 is a false track or the corresponding track of sensor 2 is misdetected while all-zero column $j$ means that track $\ell_2^{(j)}$ of sensor 2 is a false track or the corresponding track of sensor 1 is misdetedted. Conversion from the assignment (matrix) $\mathbf{S}$ to $\tau$ is given by $\tau(\ell_1^{(i)})=\sum_{j=1}^{\left|\mathbb{L}_1\right|}\ell_2^{(j)}\delta_1(S_{i,j})$.

The cost matrix of an optimal assignment problem is the $\left|\mathbb{L}_1\right|\times\left|\mathbb{L}_2\right|$ matrix:
\[
\displaystyle{ {\mathbf{C}}_{\mathbb{L}_1,\mathbb{L}_2}=\left(
\begin{array}{ccc}
{C}_{1,1}                        & \cdots         &   {C}_{1,\left|\mathbb{L}_1\right|}\\
\vdots                                  & \ddots         &   \vdots \\
{C}_{\left|\mathbb{L}_1\right|,1}     & \cdots        &   {C}_{\left|\mathbb{L}_1\right|,\left|\mathbb{L}_2\right|}
\end{array}
\right)}
\]
where for  $i\in\{1,\ldots,\left|\mathbb{L}_1\right|\}$, $j\in\{1,\ldots,\left|\mathbb{L}_2\right|\}$, ${C}_{i,j}$
is the cost of the assigning the $\ell_2^{(j)}$th track of sensor 2 to $\ell_1^{(i)}$th track of sensor 1.

 According the Definition 1, if two label spaces $\mathbb{L}_{1}$ and $\mathbb{L}_{2}$ are matched, the distance between arbitrary two single object densities (Bernoulli density) indicating the same true object from the two sensors respectively is tiny enough. RD, specifically, for $\alpha= 0.5$, equals the Hellinger affinity,  and thus the cost selection criterion becomes the equality
of Hellinger distance, which can be used to describe the distance between two  densities, which is also consistent with the LS-M criterion.  The Proposition 1 shows that single object density follows labeled Bernoulli distribution, thus using the formula for Renyi divergence between two Bernoulli distributions, it can easily be shown that
\begin{equation}\label{Cost-generalized}
\begin{split}
C_{i,j}&=\,\,\,\,\,\,R_{\alpha}(\bpi_{1}^{(\ell)}(X)||\bpi_{2}^{(\ell')}(\bX))\\&
=\frac{1}{\alpha-1}\log\int_{\bX} \bpi_{1}^{(\ell)}(\bX_n)^{(1-\alpha)} \bpi_{2}^{(\ell')}(\bX_n)^{\alpha}d\bX\\&
=\frac{1}{\alpha-1}\log\sum_{n=0}^{\infty}\frac{1}{n!}\int \bpi_{1}^{(\ell)}(\{\bx_1,\cdots,\bx_2\})^{\alpha}\\&\,\,\,\,\,\,\,\,\,\,\,\,\,\,\,\,\,\,\,\,\,\,\,\,\,\,\,\,\,\bpi_{2}^{(\ell')}(\{\bx_1,\cdots,\bx_2\})^{1-\alpha}d\bx_1\cdots d\bx_n\\&
=\frac{1}{\alpha-1}\log((1-r_{1}^{(\ell)})^{\alpha}(1-r_{2}^{(\ell')})^{1-\alpha}+\\&\,\,\,\,\,\,\,\,\,\,\,\,\,\,\,\,\,\,\,\,\,\,\,\,\,\,{r_{1}^{(\ell)}}^{\alpha}{r_{2}^{(\ell')}}^{1-\alpha}\int p_{1}^{(\ell)}(\bx)^{\alpha}p_{2}^{(\ell')}(\bx)^{1-\alpha}d\bx).
\end{split}
\end{equation}
where $\bpi_s^{(\ell)}(\bX), s=1,2$, is  the set marginal density provided in Proposition 1.

The cost of $\mathbf{S}$ is the combined costs of every true track of sensor 1 to the track of sensor 2, which can be succinctly written as
the Frobenius inner product
\[
\displaystyle{
\mathbf{J}(S)=\mbox{tr}(\mathbf{S}^{T}{C}_{\mathbb{L}_1,\mathbb{L}_2})=\sum_{i=1}^{\left|\mathbb{L}_1\right|}\sum_{j=1}^{\left|\mathbb{L}_2\right|}{C}_{i,j}{S}_{i,j}.
}
\]

The optimal assignment problem seeks an assignment matrix ${\mathbf{S}}^{*}$ ($\tau^{*}$) that minimizes the cost function $\mathbf{J}(S)$:
\begin{align}\label{Cost-1}
\begin{split}
{\mathbf{S}}^{*}=\mbox{arg} \min_{\mathbf{S}}\bJ(\mathbf{S})
\end{split}
\end{align}
where ${\mathbf{S}}^{*}$ ($\tau^{*}$) denotes the assignment matrix for the best matching hypothesis or mapping case.

By solving the equation (\ref{Cost-1}) using Murty's algorithm \cite{Murty_rank_assignment}, the true matching hypothesis $\tau^*$ is specified, then the consensual label space is given by

$$
\mathbb{L}^c=\{(\ell_1,\tau^*(\ell_1))|\tau^*(\ell_1)>0\}_{\ell_1\in\mathbb{L}_1}.
$$
 The tracks in one sensor with no corresponding matched tracks in another sensor are leaved out considering the uncertainty of them.
\begin{Rem}
The optimal $\tau^*$ establishes the optimal solution of  one-to-one matching correspondence between two label spaces, hence the consensual label space $\mathbb{L}^c$ is obtained, which makes the assumption that different sensors share the same label space come true.
\end{Rem}
\subsection{GCI Fusion with Labeled Multi-Object Density}
When fusing LMO densities via the GCI rule (\ref{G-CI}), the main challenge is that the GCI formula is computationally intractable
due to the set-integral that integrates over all joint target-spaces. However, when the condition of different label spaces matching is hold, the problem of GCI fusion with LMO densities (GCI-LMO) is great simplified, for it doesn't need to consider all possible matching correspondence between different label spaces \cite{GCI-GMB}.

In Proposition 2, we derived the GCI-fusion for generic LMO density based on the matching of label spaces.
\begin{Pro}
Let $\bpi_s(\mathbf{X})=w_s(\mathcal{L}(\cdot))P_s(\cdot)$ be the labeled multi-object posterior of sensor $s$,  $s=1,2$, and their label spaces $\mathbb{L}_1$ and $\mathbb{L}_2$ are matching. Then the distributed fusion with $\bpi_1(\mathbf{X})$ and $\bpi_2(\bX)$  via GCI rule in (\ref{G-CI}) is given by
\begin{equation}\label{GCI_LMO}
  \bpi_{\omega}(\mathbf{X})=w_{\omega}(\mathcal{L}(\mathbf{X}))p_{\omega}(\mathbf{X})
\end{equation}
where
\begin{align}\label{GCI_LMO_w}
  w_{\omega}(I)&=\frac{w_1^{\omega_1}(I)w_2^{\omega_2}(I)\eta_\omega(I)}{\sum_{I\in\mathcal{F}(\mathbb{L}_1)} w_1^{\omega_1}(I)w_2^{\omega_2}(I)\eta_\omega(I)},\\
\label{GCI_LMO_p}
p_{\omega}(\bX)&=\frac{p_1^{\omega_1}(\bX)p_2^{\omega_2}(\bX)}{\eta_\omega(I)}
\end{align}
with
\begin{align}\label{GCI_LMO_eta}
\notag \eta_\omega &(\{\ell_1,\cdots,\ell_n\})\\ =\int & p_1^{\omega_1}(\{(x_1,\ell_1),\cdots,(x_n,\ell_n)\})\\ \notag & p_2^{\omega_2}(\{(x_1,\ell_1),\cdots,(x_n,\ell_n)\})d (x_1,\cdots,x_n).
\end{align}
\end{Pro}
Especially, for special formed LMO densities (e.g., LMB \cite{delta_GLMB} and M$\delta$-GLMB \cite {Fantacci-BN} densities), their closed-form solutions had been shown in \cite{Fantacci-BT} under the assumption that different label spaces share the same birth space, and their corresponding GCI fusion is referred as GCI fusion with LMB and M$\delta$-GLMB respectively (GCI-LMB and GCI-M$\delta$-GLMB).
\subsection{Pseudo-Code}
A pseudo-code of the proposed GCI-LSM fusion algorithm is given in \textbf{Algorithm 1}.
\begin{algorithm}[h]\label{algorithm: GCI-LMB}
\caption{\label{2} {The proposed GCI-LSM fusion.}}
\textbf{Inputs}: Receive posteriors $\bpi_s$ from nodes $s=1:N_s$\;
\textbf{Step 1}: Calculate the set marginal density  of $\bpi^{(\ell)}_s$ with respect to $\bpi_s$ according to Proposition \ref{marginal_density};\\
\textbf{Step 2}: Perform  GCI-LSM fusion by adopting iteration method:\\
\textbf{Initial}: $\bpi_\omega$=$\bpi_1$\;
\For{s=2:$N_S$}
{
1) Obtain the consensual label space $\mathbb{L}^c$ of $\bpi_s$ and $\bpi_w$ according to (\ref{Cost-1})\;
2) Perform GCI-LSM fusion with $\bpi_\omega$ and $\bpi_s$ according to (\ref{GCI_LMO}), then output the fused posterior $\bpi_\omega$\;
}
\textbf{Return}: the fused posterior $\bpi_\omega(\left\{x_1,\ldots,x_n\right\})$ in the form of (\ref{factorized}).
\end{algorithm}
\section{Gaussian Mixture Implement}
We now detail the computation of the cost function $C_{i,j}$ in (\ref{Cost-generalized}) for the ranked assignment problem  of the GCI-LSM fusion for special formed LMO densities (e.g., LMB and M$\delta$-GLMB). In the present work, each single-object density conditional on its existence $p_{s}^{(\ell)}(x)$ of sensor $s$ is represented by a GM of the form
 \begin{align}
\begin{split}\label{GMs}
p_{s}^{(\ell)}(x)=\sum_{j=1}^{J_{s}^{(\ell)}}w_{s}^{(\ell,j)}\mathcal{N}\left(x;m_{s}^{(\ell,j)},P_{s}^{(\ell,j)}\right).
\end{split}
\end{align}
 Since the calculation of cost function C  involves exponentiation of GMs  which, in general, do not provide a GM. To preserve the GM form, a suitable approximation of the GM exponentiation proposed in \cite{Battistelli} is adopted. Thus, Eq. (\ref{Cost-generalized}) turns out to be
\begin{align}\label{RD}
\begin{split}
{C}&_{i,j}=-
\frac{1}{\beta}\log\left((q_1^{(i)})^{\alpha}(q_2^{(j)})^{\beta}+(r_1^{(i)})^{\alpha}(r_2^{(j)})^{\beta}K\right)
\end{split}
\end{align}
where
\begin{align}
\label{Bernoulli-K}
K&=\sum_{m=1}^{J_{1}^{(i)}}\sum_{n=1}^{J_{2}^{(j)}}w_{\omega},\\
\label{Bernoulli-w-GM}
 w_{\omega}&= \widetilde{w}_{\omega}\mathcal{N}\!\!\left(\!\!m_{1}^{(i,m)}\!-\!m_{2}^{(j,n)};0,\!\frac{P_{1}^{(i,m)}}{\omega_1}\!+\!\frac{P_{2}^{(j,n)}}{\omega_2}\!\!\right),\\
  \widetilde{w}_{\omega}&=(\!w_{1}^{(i,m)}\!)^{\omega_1}(\!w_{2}^{(j,n)}\!)^{\omega_2}\!\rho(P_{1}^{(i,m)}\!, \!\omega_1)\!\rho(P_{2}^{(j,n)}\!, \!\omega_2)
\end{align}
and  $\beta=1-\alpha$, $q_{1}^{(i)}=1-r_{1}^{(i)}$, $q_{2}^{(j)}=1-r_{2}^{(j)}$, $\rho(P,\omega)=\sqrt{\det[2\pi P\omega^{-1}](\det[2\pi P])^{-\omega}}$.

Moreover, the GM implementation of the GCI fusion for special formed LMO densities (e.g., LMB and M$\delta$-GLMB) under the assumption that different sensors share the same label space can refer to (55)  in \cite{Battistelli}.
\section{Performance Assessment}
The performance of the proposed GCI-LSM fusion is evaluated  in two 2-dimensional multi-object tracking scenarios. The GCI-LSM is implemented using the  GM approach proposed in Section V. Since this paper does not focus on the problem of weight selection, we choose the Metropolis weights for convenience (notice that this may have an impact on the fusion performance). The LMB  filter is adopted by local filters.  The efficiency of LMB filter has been demonstrated in \cite{delta_GLMB}.

All targets travel in straight paths and with different but constant velocities. The number of targets is time varying due to births and deaths. The following target and observation models are used. The target state variable is a vector of plannar position and velocity $x_k=[p_{x,k},\dot{p}_{x,k},p_{y,k},\dot{p}_{y,k}]^{\top}$, where ``$^\top$'' denotes the matrix transpose.
The single-target transition model is linear Gaussian specified by
\[
\displaystyle{
 \bF_k=\left[
\begin{array}{cc}
\bI_2& \Delta\bI_2 \\ \mathbf{0}_2 & \bI_2
\end{array}
\right],\,\,\,\,\bQ_k=\sigma_v^{2}\left[
\begin{array}{cc}
\frac{1}{4}\bI_2& \frac{1}{2}\Delta\bI_2 \\ \frac{1}{3}\mathbf{0}_2 & \bI_2
\end{array}
\right]
}
\]
where $\bI_n$ and $\mathbf{0}_n$ denote the $n\times n$ identity and zero matrices, $\Delta=1$ second (s) is the sampling period, and $\sigma_\nu=5m/s^{2}$ is the standard deviation
of the process noise. The probability of target survival is $P_{S,k}=0.99$; The probability of target detection in each sensor
is independent of the probability of detection at all sensors and is $P_D=0.99$. The single-target observation model is also a linear Gaussian with
\[
\displaystyle{
 \bH_k=\left[
\begin{array}{cc}
\bI_2  & \mathbf{0}_2
\end{array}
\right],\,\,\,\,
\mathbf{R}_k=\sigma_\varepsilon^{2}\bI_2,
}
\]
where $\sigma_\varepsilon=1.4m$, is the standard deviation of the measurement noise. The number of clutter reports
in each scan is Poisson distributed with $\lambda=10$. Each clutter report is sampled uniformly over the whole surveillance region.

The parameters of GM implementation have chosen as follows: the truncation threshold is  $\gamma_t=10^{-4}$; the prune  threshold is $\gamma_p=10^{-5}$; the merging threshold is $\gamma_m=4$; the maximum number of Gaussian components is $N_{max}=10$. All performance metrics are given in term of the optimal sub-pattern assignment (OSPA) error \cite{MeMBer_Vo1}.
\subsection{Scenario 1}
\begin{figure}[h]
\centering
\includegraphics[width=8cm]{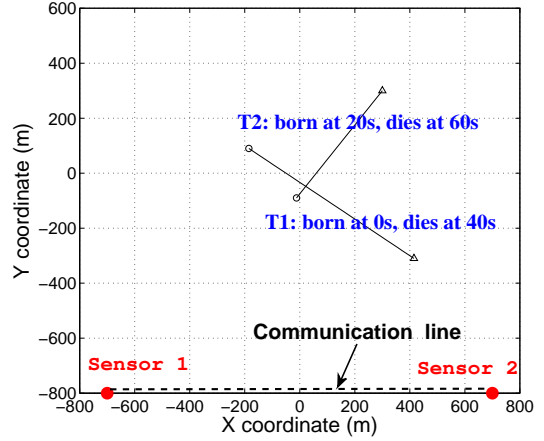}
\caption{{\bf } The scenario of simple distributed sensor network with two sensors tracking two targets.}
\label{fig_sim}
\end{figure}
To demonstrate the effectiveness of GCI-LSM fusion, the performance of GCI-LSM fusion is compared with GCI-LMB \cite{Fantacci-BT} (under the assumption that different sensors share the same label space) in two experiments with ABP and PBP used respectively
 For this purpose, an simple scenario involving two sensors and two objects is considered as shown in Fig 2. The duration of this scenario is $T_s=60s$.
\subsubsection{Experiment 1}
 The preceding analyses show that the LS-DM phenomenon arises frequently when the ABP is adopted by local filters. To prove this and the effectiveness of the proposed GCI-LSM fusion, the GCI-LSM fusion is compared with GCI-LMB fusion under the ABP situation.

 The adaptive birth procedure proposed in \cite{delta_GLMB} is employed for this scenario. More specifically, the existence probability of the Bernoulli birth distribution at time $k+1$ depending on the measurement $z_k$ is proportional to the probability that $z_k$ is not assigned to any target during the updated at time $k$:
\begin{align}\label{adaptive_birth}
r_{B,k+1}(z)=\min\left(r_{B,\max},\frac{1-r_{U,k}(z)}{\sum_{\xi\in Z_k}1-r_{U,k}(\xi)}\cdot \lambda_{B,k+1}\right)
\end{align}
where
\begin{align}\label{adaptive_birth_r_1}
r_{U,k}=\sum_{(I_+,\theta)\in \mathcal{F}(\mathbb{L}_+)\times \Theta_{I_+}}1_{\theta}(z) w_k^{I_+,\theta}
\end{align}
with $w_k^{I_+,\theta}$ is given by (59) in \cite{delta_GLMB}, and $\lambda_{B,k+1}$ is the expected number of target birth at time $k+1$ and $r_{B,\max}\in[0, 1]$ is the maximum existence probability of a new born target.

 \begin{figure}[h]
\centering
\includegraphics[width=5cm]{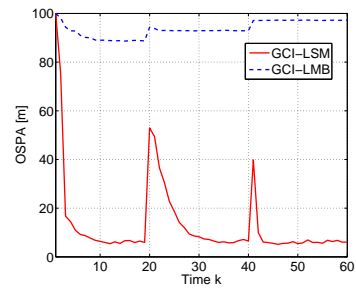}
\caption{{\bf } ABP: OSPA errors of GCI-LSM and GCI-LMB fusion algorithms with order $p=1$ and cut-off  $c=100$ with adaptive birth (200 MC runs).}
\label{fig_sim}
\end{figure}
Fig. 3 illustrates that the performance of GCI-LSM fusion is significant better than GCI-LMB fusion. Since the method of the ABP  depends on the observations with randomness, the labels of the birth targets each time are also randomness with their corresponding observations.  This result of GCI-LMB fusion shows that the ABP leads to the LS-DM frequently, and it is necessity to match the label spaces from different sensors to ensure the  consensual between them, or the performance of  the GCI fusion will collapse. The result of GCI-LSM also evidences this viewpoint. It can seen that once the LS-DM is removed, the GCI-fusion perform really excellent, exactly as the GCI-LSM fusion. The outstanding performance of GCI-LSM   also gets benefit from the well-designed cost function in ranked assignment problems.
\subsubsection{Experiment 2}
 This experiment   analyzes the problem of the LS-DM under PBP situation. The preceding analyses  show that the PBP also suffers from the LS-DM even though it can obtain some priors about the births. In this experiment, the performance of GCI-LSM and GCI-LMB fusion is compared using PBP \cite{delta_GLMB}.
\begin{figure}[htbp]
\begin{minipage}[htbp]{0.47\linewidth}
  \centering
  \centerline{\includegraphics[width=5cm]{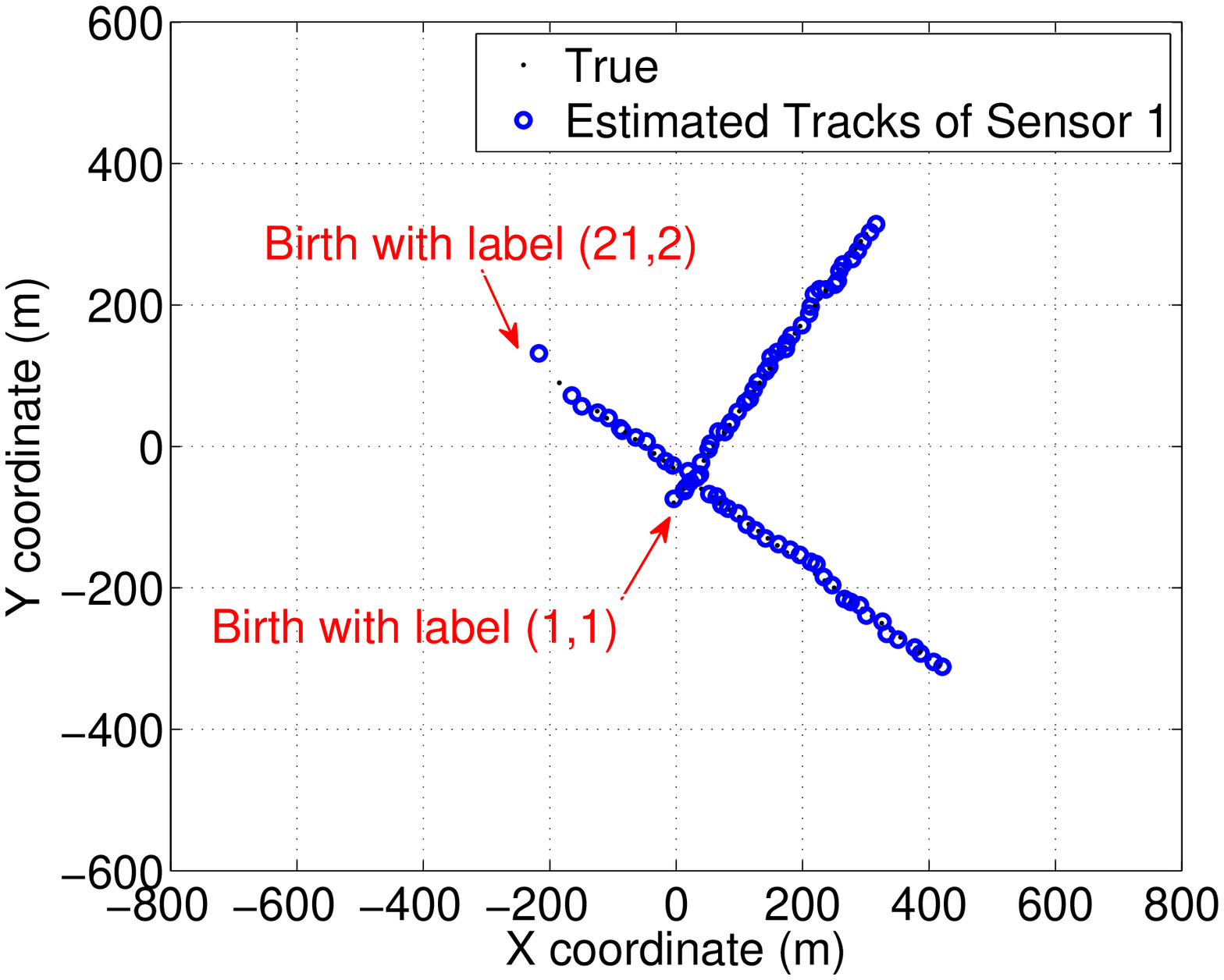}}
  \centerline{\small{\small{(a)}}}\medskip
  \end{minipage}
  \hfill
\begin{minipage}[htbp]{0.47\linewidth}
  \centering
  \centerline{\includegraphics[width=5cm]{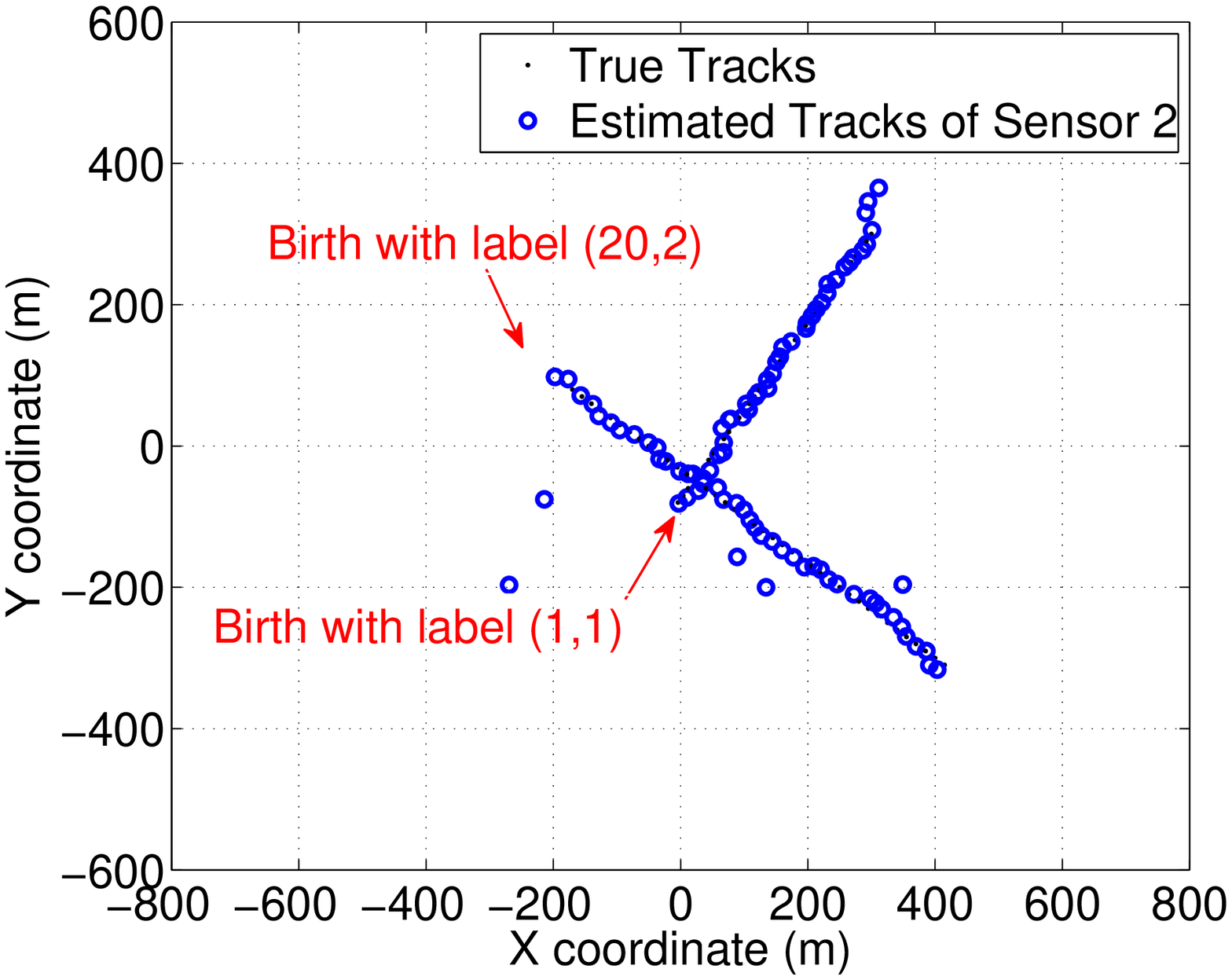}}
  \centerline{\small{\small{(b)}}}\medskip
\end{minipage}
  \hfill
\begin{minipage}[htbp]{0.47\linewidth}
  \centering
  \centerline{\includegraphics[width=5cm]{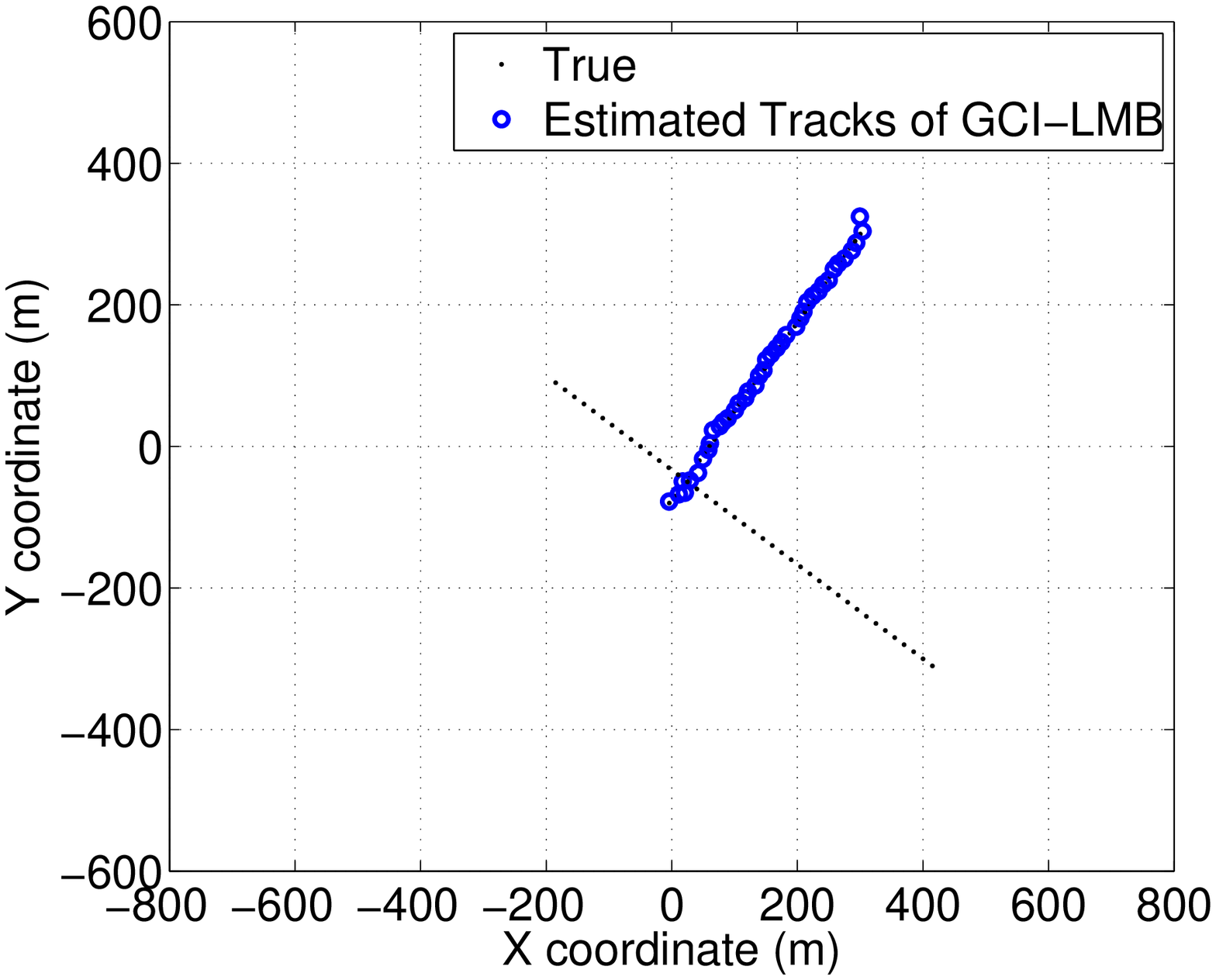}}
  \centerline{\small{\small{(c)}}}\medskip
\end{minipage}
  \hfill
\begin{minipage}[htbp]{0.47\linewidth}
  \centering
  \centerline{\includegraphics[width=5cm]{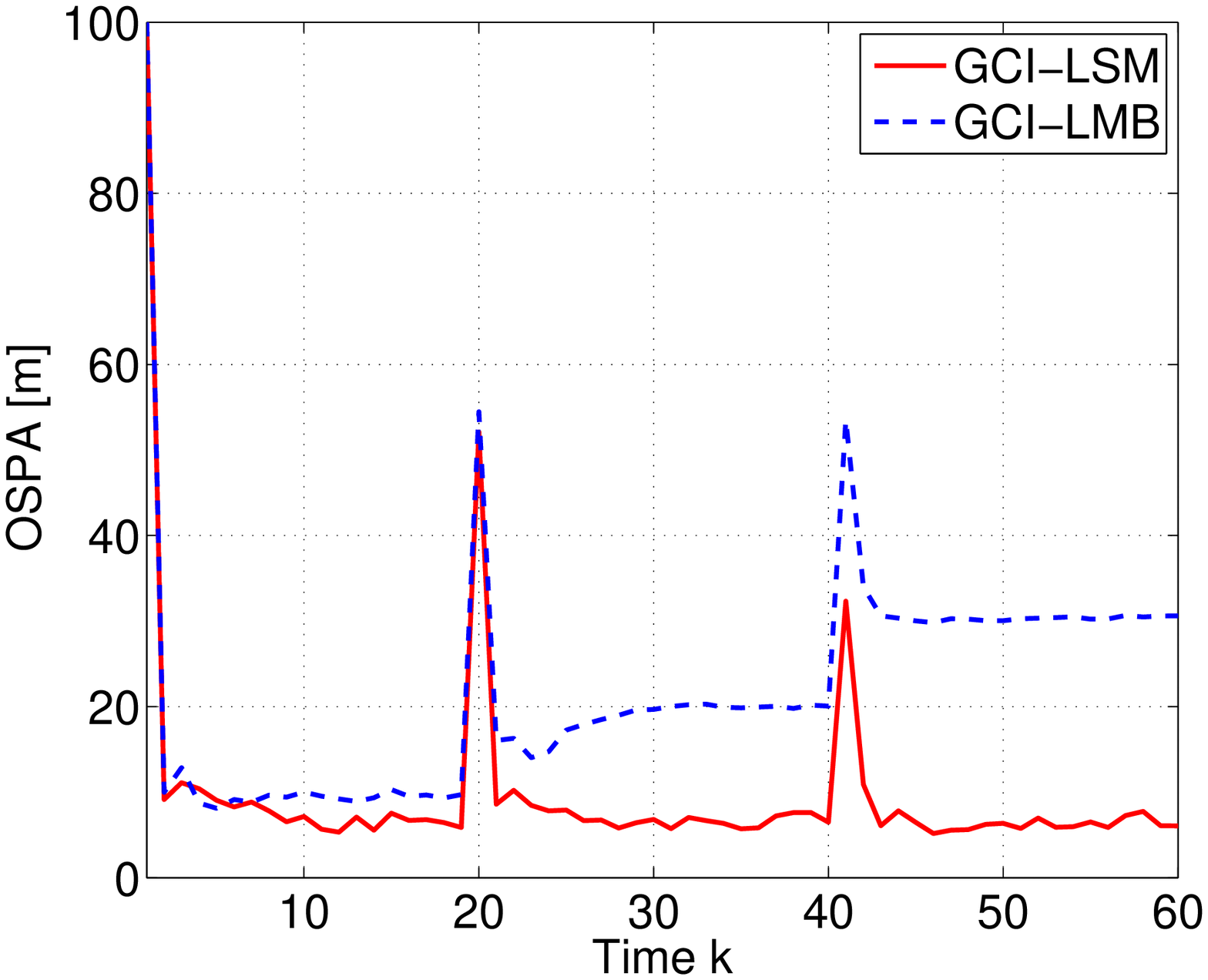}}
  \centerline{\small{\small{(d)}}}\medskip
\end{minipage}
\caption{PBP procedure: (a) multi-object state estimation of sensor 1 (single MC run), (b) multi-object state estimation of sensor 2 (single MC run), (c) multi-object state estimation of GCI-LMB fusion (single MC run),(d) OSPA errors of GCI-LSM and GCI-LMB fusion algorithms  with  order $p=1$ and cut-off  $c=100$ (200 MC runs).}
\label{measure_based_mismatching_phenomenon}
\end{figure}

Figs. 4 (a)-(c) show the multi-object estimations of local filters and GCI-LMB fusion respectively for single MC run. It can be seen that in this run, the GCI-LMB fusion fails to perform fusing for the $2$th track while both local filters accurately estimate this track.  Due to that the prior information for births only provide the initial positions, but fail to provide the initial time, $2$th object is initialized at different time step  in different local filters. Hence, the labels of the $2$th object of different local filters are mismatching obviously, leading to that the GCI-LMB fusion algorithm completely lose the $2$th object. The performance comparison between GCI-LSM fusion and GCI-LMB is also shown in Fig. 4 (d). As expected, the performance of the GCI-LSM fusion has remarkable advantages towards GCI-LMB fusion, and  the GCI-LMB fusion is getting worse with target births and deaths. This result is consistent with the single Monte Carlo (MC) run's. The above results confirm that the GCI-LSM fusion is able to handle the LS-DM, while the GCI-LMB fusion cannot.
\subsection{Scenario 2}
\begin{figure}[h]
\centering
\includegraphics[width=8cm]{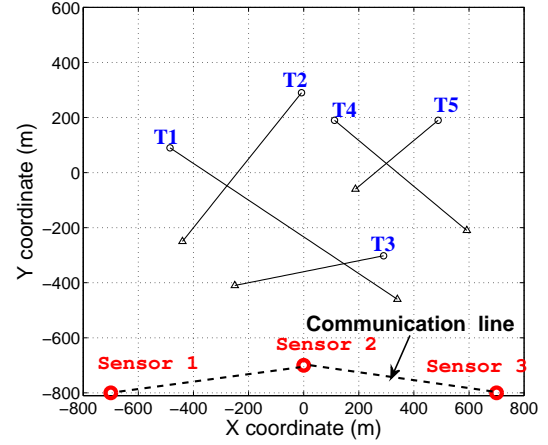}
\caption{{\bf } The scenario of distributed sensor network with three sensors tracking five targets. T1 born at 0s, dies at 55s; T2 born at 0s, dies at 55s; T3 born at 10s, dies at 65s, T4 born at 25s, dies at 65s; T5 born at 40s, dies at 65s.}
\label{fig_sim}
\end{figure}
To further test the performance of the proposed GCI-LSM fusion in challenging scenarios, a sensor network scenario involving five targets is considered as shown in Fig. 5.  In the experiment, the proposed GCI-LSM fusion is compared to the  GCI-GMB fusion mentioned in Section III-B \cite{GCI-GMB} and the GCI fusion with PHD filter (GCI-PHD)\cite{Uney-2}. Both GCI-LSM and GCI-GMB fusion use ABP introduced in scenario 1 and the adaptive birth distribution of GCI-PHD is introduced in \cite{Ristic_PHD}.  The duration of this scenario is $T_s=65s$.
\begin{figure}[htbp]
\begin{minipage}[htbp]{0.47\linewidth}
  \centering
  \centerline{\includegraphics[width=5cm]{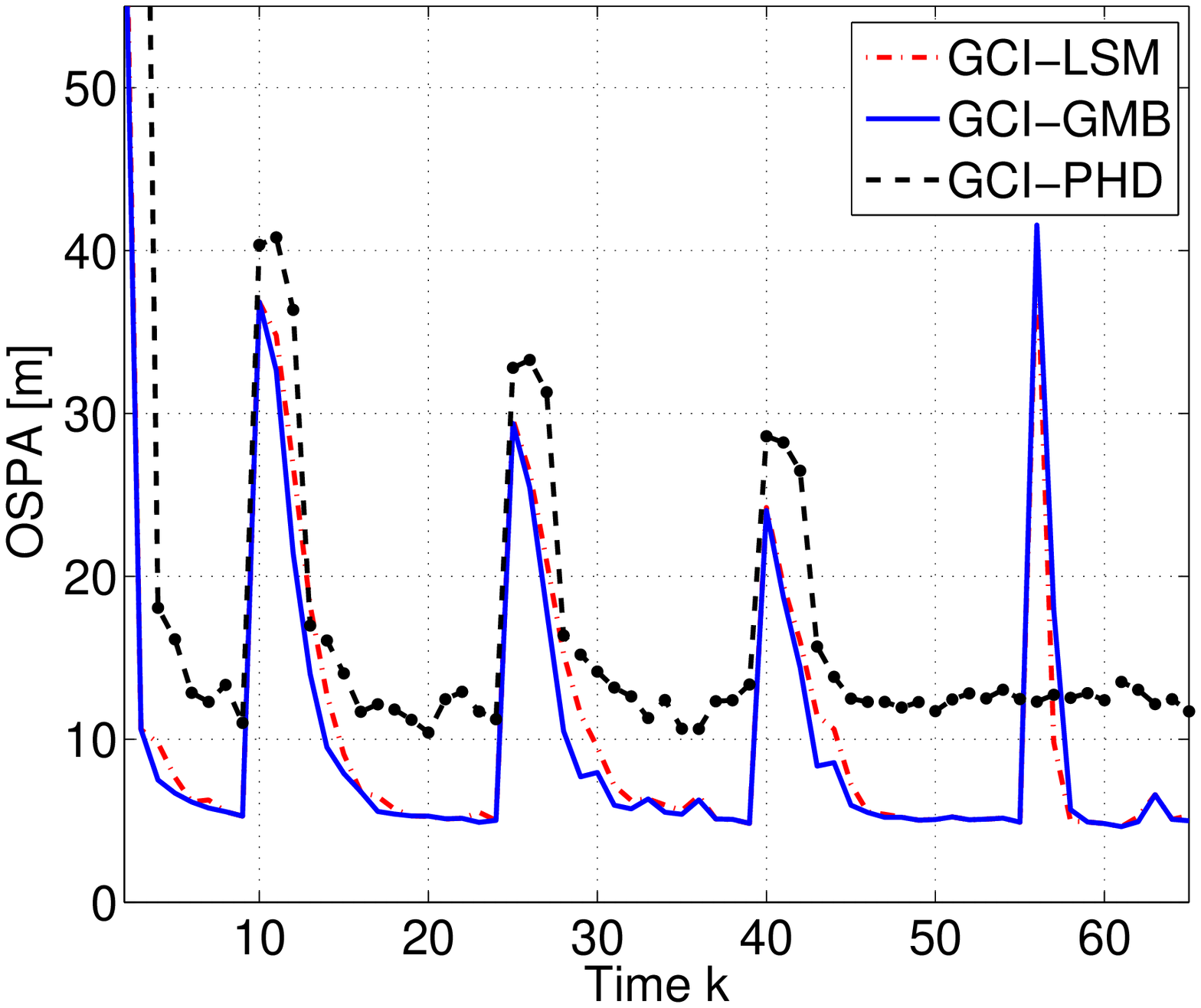}}
  \centerline{\small{\small{(a)}}}\medskip
  \end{minipage}
  \hfill
\begin{minipage}[htbp]{0.47\linewidth}
  \centering
  \centerline{\includegraphics[width=5cm]{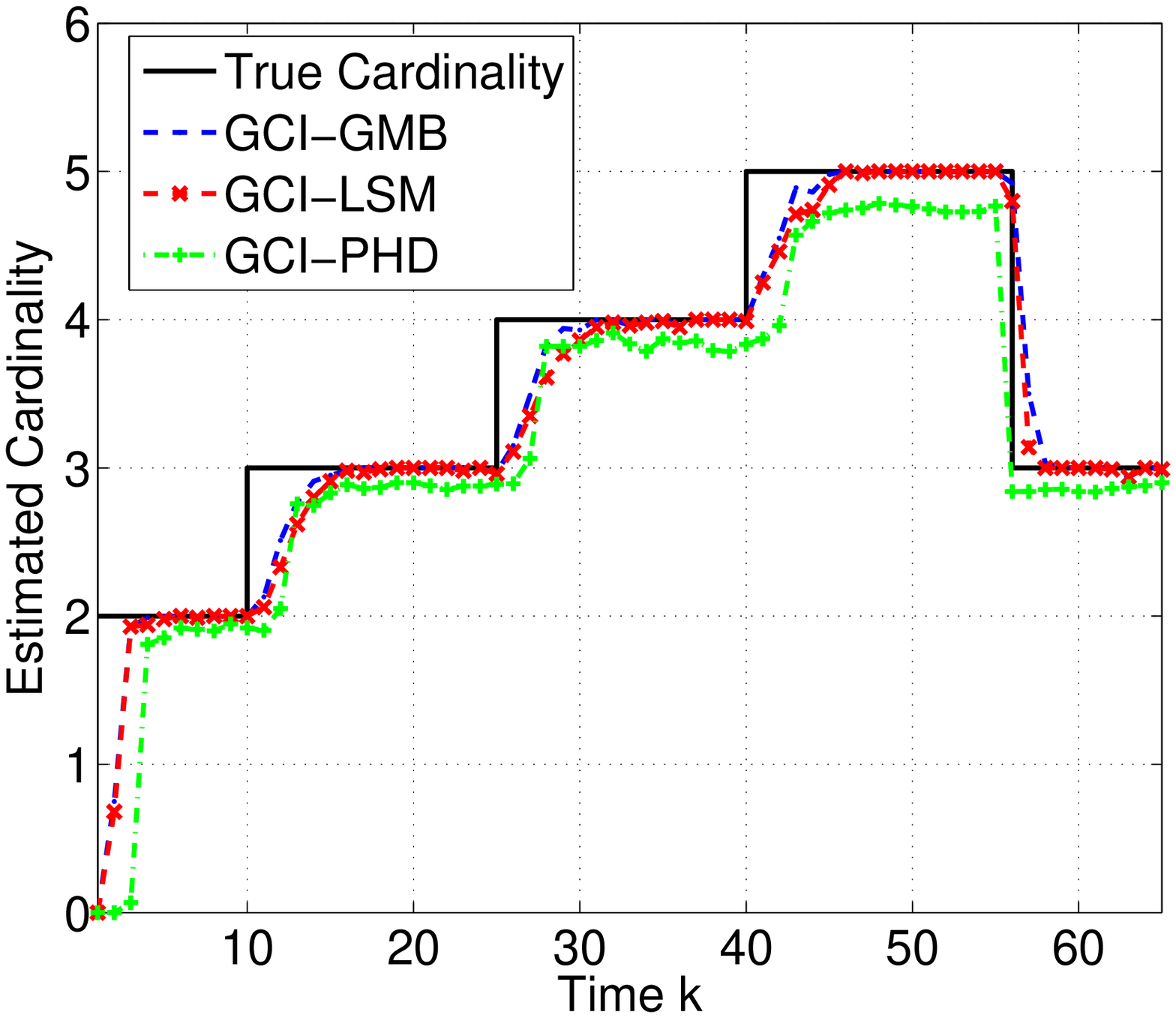}}
  \centerline{\small{\small{(b)}}}\medskip
\end{minipage}
\caption{(a) OSPA distance of order $p=1$ and cut-off  $c=100$ for the GM implementation with adaptive birth, (b) cardinality estimation, (200 MC
runs).}
\label{measure_based_mismatching_phenomenon}
\end{figure}

Both the OSPA distance and the cardinality estimation in Fig. 6 illustrate the performance differences among the three fusion methods. It can be seen that the performance of GCI-LSM is almost the same as GCI-GMB  after the performances converge.   Also the GCI-LSM performs slightly worse than GCI-GMB when objects are born, and the explanation here is that GCI-GMB fusion considered all possible matching correspondences between label spaces of different sensors jointly, while GCI-LSM  fusion utilizes an optimal estimation of the matching correspondence. Moreover the tiny performance loss of GCI-LMS fusion toward GCI-GMB fusion also demonstrates the superiority  of the  optimal estimation of matching correspondence. In other words, the ranked assignment problem built in Section IV-B  can match the label spaces from different sensors accurately and consistently. In addition, Fig. 6 also reveals that both the GCI-LSM and GCI-GMB fusion outperform the GCI-PHD fusion for both OSPA error and the cardinality.  This result also demonstrates the effectiveness  GCI-LSM fusion.


\section{Conclusion}
This paper investigates the problem of distributed multi-target tracking (DMMT) with labeled multi-object density based on generalized covariance intersection.
Firstly, we  provided a principled mathematical definition of label spaces matching (LS-M) based on information divergence,  referred to as LS-M criterion.  Then we proposed a novel two-step distributed fusion algorithm. Firstly, to match the label spaces from different sensors, we   build  a ranked assignment problem to seek the optimal solution of matching correspondence between objects of different sensors based on LS-M criterion. Then,  GCI fusion is performed on the matched label space. Moreover, we derive the GCI fusion with generic labeled multi-object (LMO) densities.
A Gaussian mixture implementation of the proposed GCI-LSM is also given, and its effectiveness and better performance are demonstrated in numerical results.  At the present stage, the impact of objects closely spaced on the fusion is not very clearly, thus  further work will study the   GCI-LSM fusion  considering objects in proximity.




\end{document}